\documentclass{article}
\usepackage{arxiv}
\usepackage[utf8]{inputenc} 
\usepackage[T1]{fontenc}    
\usepackage{hyperref}       

\usepackage{url}            
\usepackage{booktabs}       
\usepackage{amsfonts}       
\usepackage{nicefrac}       
\usepackage{microtype}      
\usepackage{lipsum}
\usepackage{graphicx}

\usepackage{amsmath}
\usepackage{amssymb}
\usepackage{mathtools}
\usepackage{amsthm}
\usepackage[capitalize,noabbrev]{cleveref}
\usepackage{makecell}
\usepackage{subfigure}

\title{Alleviating Hyperparameter-Tuning Burden in SVM Classifiers for Pulmonary Nodules Diagnosis with Multi-Task Bayesian Optimization}

\author{
 Wenhao Chi \\
  Yau Mathematical Sciences Center\\
  Tsinghua University\\
  Beijing, China\\
  \texttt{whchi@mail.tsinghua.edu.cn} \\
   \And
 Haiping Liu \\
  PET/CT Center\\
  The First Affiliated Hospital of Guangzhou Medical University \&\\
  China State Key Laboratory of Respiratory Disease\\
  Guangzhou, China\\
  \texttt{lhp586@163.com} \\
  \And
  Hongqiao Dong\\
  Academy of Mathematics and Systems Science\\
  Chinese Academy of Sciences \&\\
  University of Chinese Academy of Sciences\\
  Beijing, China\\
  \texttt{hqdong@amss.ac.cn}\\
  \And
  Wenhua Liang\\
  Department of Thoracic Surgery and Oncology\\
  The First Affiliated Hospital of Guangzhou Medical University \&\\
  China State Key Laboratory of Respiratory Disease\\
  Guangzhou, China\\
  \texttt{liangwh1987@163.com}\\
  \And
  Bo Liu\\
  Academy of Mathematics and Systems Science\\
  Chinese Academy of Sciences\\
  Beijing, China\\
  \texttt{bliu@amss.ac.cn}
}

\begin{document}
\maketitle
\begin{abstract}
In the field of non-invasive medical imaging, radiomic features are utilized to measure tumor characteristics. However, these features can be affected by the techniques used to discretize the images, ultimately impacting the accuracy of diagnosis. To investigate the influence of various image discretization methods on diagnosis, it is common practice to evaluate multiple discretization strategies individually. This approach often leads to redundant and time-consuming tasks such as training predictive models and fine-tuning hyperparameters separately. This study examines the feasibility of employing multi-task Bayesian optimization to accelerate the hyperparameters search for classifying benign and malignant pulmonary nodules using RBF SVM. Our findings suggest that multi-task Bayesian optimization significantly accelerates the search for hyperparameters in comparison to a single-task approach. To the best of our knowledge, this is the first investigation to utilize multi-task Bayesian optimization in a critical medical context.
\end{abstract}


\section{Introduction}

Against the backdrop that lung cancer continues to be the leading cause of cancer mortality worldwide, precisely differentiation between benign and malignant pulmonary nodules has always been the focus of machine learning and clinical medicine \cite{travis2013new, li2019performance, chatterjeeadvancement}. Machine learning based pulmonary nodule diagnosis has opened up new opportunities to relax the limitation from physicians’ subjectivity, experiences and fatigue, thereby speeding up the diagnostic process and saving more lives \cite{liu2020}.

Distinguishing malignant and benign nodules traditionally involves tasks such as image preprocessing, feature learning and classification model construction, which affect each other \cite{katre2017}. For instance, the radiomic features extracted in the feature learning are easily affected by the image discretization strategy in the image preprocessing, which in turn affects the diagnostic accuracy of the classification model \cite{van2016repeatability}. Therefore, to achieve better diagnostic accuracy, it is necessary to find a set of optimal policy combinations from the policy space which defines the approaches and/or their parameters for implementing each task.

In reality, related work is limited to only finding combinations of strategies and parameters for a specific task without explicitly acknowledging inter-task relationships. Ignorance of inter-task relationships frequently induces repetitive algorithm-tuning work that is dependent on expert knowledge, and impacts of strategies and parameters used in this task on subsequent tasks are unpredictable. Furthermore, it induces huge overhead accompanied by time-expensive medical image processing.

To quickly evaluate the performance of the diagnostic model under different policy combinations, alleviating the burden of model hyperparameter tuning, we, consequently, charted an alternative route by leveraging multi-task Bayesian optimization (MTBO) \cite{swersky2013,shahriari2016}. By appreciating the useful inter-task relationships, an accelerated search could be expected by transferring these commonalities among tasks, and thereby possibly alleviate time-consuming repeated searches. On the way to this destination, we raise two challenging questions:

\textit{Can multiple medical classification tasks be solved simultaneously by sharing knowledge across tasks, rather than optimizing one by one?}

\textit{By using multi-task Bayesian optimization, are speed-ups in search and less loss achievable?}

\begin{figure*}[htbp]
\begin{center}
\centerline{\includegraphics[width=\textwidth]{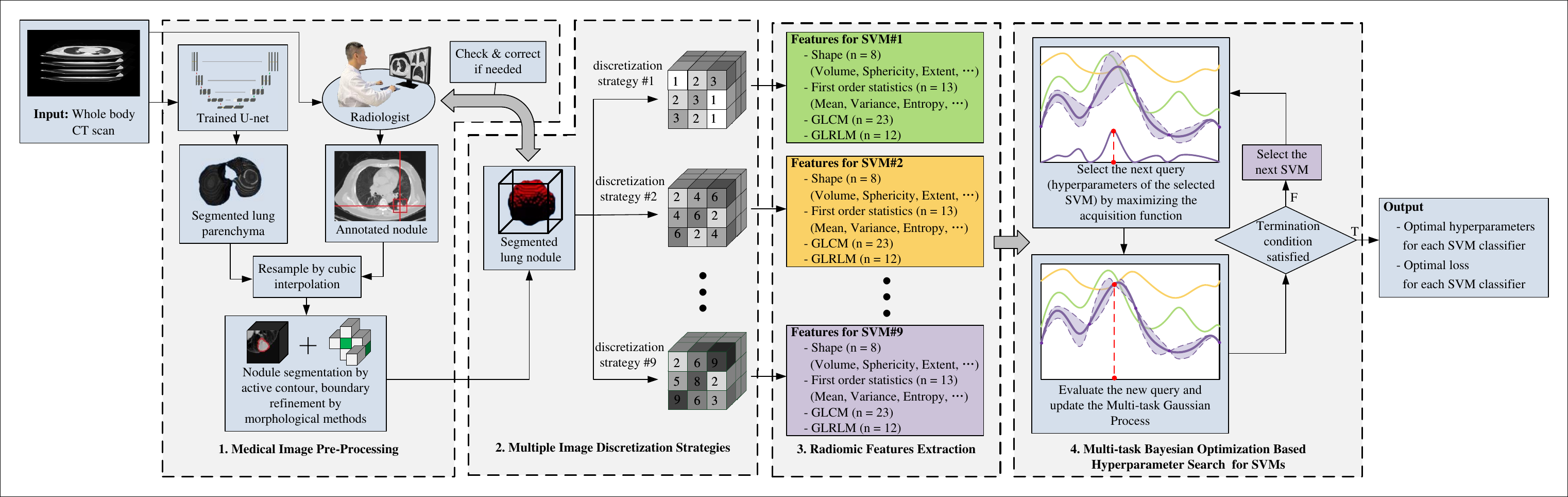}}
\caption{\textbf{Workflow of the proposed MTBO based SVMs hyperparameter-tuning method on pulmonary nodule classification}. The collected whole body CT scans are first preprocessed to delineate the pulmonary nodules. Then, the segmented nodules are discretized via multiple strategies. After that, for the result of each discretization strategy, various types of features are extracted as the input to a specific SVM. Finally, a multi-task Gaussian process is established through the correlation between tasks, the query point is selected by maximizing the acquisition function, and the optimal hyperparameters are determined by iterative query.}
\label{fig:workflow}
\end{center}
\end{figure*}

Our contributions are two-fold, and \cref{fig:workflow} summarizes the complete flow of this study.

For the first question, we generate multiple medical tasks (distinguishing benign from malignant pulmonary nodules) through different image discretization strategies. Then the RBF SVMs are used as the classification models, and the hyperparameters of multiple SVM classifiers are tuned simultaneously using MTBO to obtain the optimal model. This is the first study to apply MTBO technology to the medical field.

For the second question, the proposed algorithm \textit{MTBO for multiple SVM classifiers} efficiently finds suitable hyperparameters for the diagnostic models and generally outperforms the performance of single-task Bayesian optimization. Moreover, in the optimization of multi-tasks that require a lot of time to evaluate, MTBO has greater application potential.

The application of machine learning in medical imaging still faces daunting challenges. Our research aims to narrow the gap between machine learning research and clinics-oriented decision-making practice.

\section{Related Work}
\label{Related Work}
\textbf{Image discretization}, resampling of voxel intensities within the ROI to a limited range of gray values (also known as bin), is a necessary procedure during the medical image analysis \cite{Yip2016} Various image discretization strategies can be generated by changing the bin width, the bin number, or the quantization range, but there is currently no unified standard. Studies have shown that image discretization strategies can substantially influence the stability of radiomic features \cite{leijenaar2013,larue2017,shafiq2018,park2020}. No relevant study has investigated whether different discretization strategies directly affect the diagnosis of benign-malignant pulmonary nodules.

\textbf{Pulmonary Nodule classification} differentiates between benign and malignant nodules on different image formats \cite{liu2020}. The general workflow of handling a pulmonary nodule classification task based on traditional machine learning methods includes (1) Preprocessing images to make them uniform in format and easy to compute \cite{kulkarni2014}; (2) Extracting features that can describe the nature of the images or nodules (shape, statistic, texture, etc.) as the input to the classification models \cite{hawkins2014,junior2018,zhou2018}; (3) Building models(LASSO, SVM, random forest, neural networks, etc.) to make the final judgment \cite{touw2013,shi2005,sun2013}. The convolutional neural network(CNN) is another widely used tool in pulmonary nodule classification, which can automatically learn the implicit information in images without the need for manual feature extraction \cite{yuan2018,shen2019,xie2018}.
During the whole process, people need to make a trade-off between the complexity of features and models and their generalization performance, that is, simple features or models can hardly express all the information that determines the nodule’s malignancy, while complicated ones may lead to overfitting, long training time, and oversensitivity to hyperparameters. Therefore, choosing the most suitable method among the various feasible options becomes the key to solving the problem. However, trying to compare all the combinations together is extremely time-consuming. To the best of our knowledge, there are no studies on how to improve efficiency in evaluating multiple models in the field of lung pulmonary classification.

\textbf{Hyperparameter-tuning} is an essential part of training an efficient machine learning model. For small-scale problems and simple-structured models, grid search is the most widely used strategy for hyperparameter-tuning. For instance, when building a survival model based on the LASSO method, researchers usually select several penalty coefficients at equal intervals and use cross-validation to determine the optimal value \cite{gevaert2012,huang2016}. For complex models (such as deep learning) or large datasets, function evaluation can be very expensive, random search \cite{Bergstra2012} and Bayesian optimization \cite{Snoek2012} will be more appropriate than exhaustive algorithm. In the medical field, Bayesian optimization has been successfully applied to automatic parameter tuning of deep learning \cite{Borgli2019}. In multi-task scenarios, the hyperparameter-tuning burden is heavier. Swersky et al. \cite{swersky2013} proposed multi-task Bayesian optimization to exploit the similarity between tasks to improve the overall optimization efficiency. However, the technique of multi-task Bayesian optimization has not been applied to multiple medical tasks.

\textbf{Multi-task learning} is a machine learning paradigm aims to improve the generalization performance of multiple related learning tasks by lever-aging domain-specific information between tasks \cite{caruana1997}. The paradigm has been applied in cancer research to improve predictive performance \cite{gonen2014,khosravan2018}. Inspired by the concept of multi-task learning, multi-task optimization has been developed. In machine learning, optimizing the tasks by sharing features between related datasets can improve the efficiency of the training process \cite{chandra2016}, but multi-task optimization has little application in the medical field.

\section{Proposed Method}
\label{Proposed Method}
In this section, we introduce nine image discretization strategies and multi-task Bayesian optimization, and then propose an algorithm to guide how to use multi-task Bayesian optimization to tune parameters for multiple SVMs simultaneously. 

\subsection{Image Discretization Strategies}

In medical image analysis, image discretization groups intensity values within the region of interest into smaller bins, which is the premise of radiomic feature extraction \cite{Yip2016}. The intensity $I_o(v_k)$  of voxel $v_k$  in original image is transformed to a discretized intensity $I_d(v_k)$  by following \cref{eq:q,eq:I}.
\begin{align}\label{eq:q}
    q_l = q_0 + \frac{(q_N-q_0)}{N_g}l \quad 1 \le l \le N_g-1, 
\end{align}
\begin{align}\label{eq:I}
    I_d(v_k) = \left\{ \begin{array}{rl}
    1,   & \text{if }l_o(v_k)\le q_1 \\
    l,   & \text{if } q_{l-1}<l_o(v_k)\le q_l \\
    N_g,   & \text{if }l_o(v_k)> q_{N_g-1}    
    \end{array}\right.,
\end{align}
where $N_g$  denotes number of bins, and $\{q_l\}_{l=0}^{N_g}$ defines a sequence used as the reference to discretize the original voxel’s intensity.
Obviously discretization highly depends on user-specified triplet $\{N_g,q_0,q_{N_g}\}$ that controls the sensitivity of the discretization to noise. Our primary concern is how best to eliminate the negative impact of inappropriate discretization strategies on the followup procedures (i.e. radiomic feature computation and classification). 

In this study, we designed multiple discretization strategies, that is, different combinations on $\{N,q_0,q_{N_g}\}$. We chose three bin numbers (16, 32, and 64) and three quantization ranges (min-max, mean ± 2SD, and mean ± 3SD). Nine discretization strategies were generated in total. Instead of evaluating one by one, we adopted the multi-task Bayesian optimization to evaluate them simultaneously to save computational costs. Next, we introduce the multi-task Bayesian optimization.

\subsection{Multi-Task Bayesian Optimization}

Multi-task Bayesian optimization leverages the commonality between tasks to facilitate time-saving learning \cite{swersky2013, shahriari2016}. The expensive-to-evaluate function is represented by a cheap multi-task Gaussian surrogate model updated by feeding new observations. To lighten the burden of acquiring new observations, the inter-task dependence guides selection of promising points. This dependence is learned by a positive semi-definite covariance matrix over tasks. It sped up the search for hyperparameters. This paper explores the possibility of using multi-task Bayesian optimization to accelerate the search for hyperparameters in multiple SVM classifiers for benign–malignant classification of pulmonary nodules.

In multi-task Bayesian optimization, supposed that there are $M$ related tasks, given a set $\mathbf{X}$ of $N$ distinct points $\mathbf{X}=\{\mathbf{x}_i\}_{i=1}^N \subset \mathcal{X} \subset \mathbb{R}^d $ we define $y_{ij}$ as the response of the $i^{th}$ task under $\mathbf{x}_i$, and $\mathbf{Y} = (y_{11},\dots,y_{1N},y_{21},\dots,y_{2N},\dots,y_{M1},\dots,y_{MN})^T$. Multi-task Gaussian process \cite{Bonilla2007} regression is an appropriate choice to find correspondence between $\mathbf{x}$ and $\{\mathbf{y}_i\}_{i=1}^M$. For $\forall \mathbf{x} \in \mathcal{X}$, $\forall i\in\mathcal{T}=\{1,2,\dots,M\}$, define a Gaussian process with $(\mathbf{x},i)$ as input, i.e., $f:\mathcal{X}\times\mathcal{T}\to \mathbb{R}$
\begin{align}
    f(\mathbf{x},i)&\sim \mathcal{GP}(\mu,k),\\
    y &\sim \mathcal{N}(f(\mathbf{x},i),\sigma_i^2),
\end{align}
where $\mu:\mathcal{X}\times\mathcal{T}\to \mathbb{R}$ and $k:\mathcal{X}\times\mathcal{T}\times\mathcal{X}\times\mathcal{T}\to \mathbb{R}$ are Gaussian process’ prior mean function and covariance function, respectively, $\sigma_i^2$ is the noise variance for the $i^th$ task. We set 
\begin{align}
    \mu(x,i) &= \mu_i,\\
    k(\mathbf{x},i,\mathbf{x}',l) &= K_{il}^tK^x(\mathbf{x},\mathbf{x}'),
\end{align}
where $\mathbf{\mu}=(\mu_1,\mu_2,\dots,\mu_M)^T$ is a constant vector, $K^t$ is a positive semi-definite matrix that measures the relationship between tasks, and $K^x$ is a covariance function that measures the relationship between points on $\mathcal{X}$. We choose Matérn 5/2 covariance function for $K^x$. Formally,
\begin{align}
    K^x(\mathbf{x},\mathbf{x}') &= \sigma_f^2\left(1+\frac{\sqrt{5}r}{\sigma_l} + \frac{5r^2}{3\sigma_l^2} \right)\exp\left(-\frac{\sqrt{5}r}{\sigma_l} \right),\\
    r &= \sqrt{(\mathbf{x} - \mathbf{x}')^T(\mathbf{x} - \mathbf{x}')},
\end{align}
without loss of generality, we can set $\sigma_f=1$ because the variance can be fully explained by $K^t$.
Then the predictive mean and covariance of the new point $\mathbf{x}$ on the $i^{th}$ task can be inferred as
\begin{align}
    \mu(\mathbf{x},i;\mathbf{X},\mathbf{Y},\theta) = \mu_i (K_{\cdot i}^t\otimes K^x(\mathbf{X},\mathbf{x}))^T\Sigma^{-1}_0(\mathbf{Y}-\mathbf{\mu}\otimes I_N)
\end{align}
\begin{align}
    \Sigma(\mathbf{x},i, \mathbf{x}',l;\mathbf{X},\mathbf{Y},\theta) = K_{il}^tK^x(\mathbf{x},\mathbf{x}') - (K_{\cdot i}^t\otimes K^x(\mathbf{X},\mathbf{x}))^T\Sigma^{-1}_0(K_{\cdot l}^t\otimes K^x(\mathbf{X},\mathbf{x}')),
\end{align}
where $\otimes$ denotes the Kronecker product, $\theta$ represents all parameters, which will be estimated by maximum likelihood, $K_{\cdot i}$ represents the $i^th$ column of $K^t$, $\Sigma_0 = K^t\otimes K^x(\mathbf{X},\mathbf{X}) + D\otimes I_N$, $D=\text{diag}(\sigma_1^2,\dots,\sigma_M^2)$, $I_N$ is an $N$-dim identity matrix. If there are missing values in $\mathbf{Y}$, we impute them with the prior means.

To find the minimum value of $f$, it is usually to maximize an acquisition function $a(\mathbf{x},i;\mathbf{X},\mathbf{Y},\theta)$ to determine the next query point $(\mathbf{x}^{next},i^{next})$. We use the expected improvement (EI) criterion, let $y_{best} = \min{\mathbf{Y}}$ be the currently observed minimum, $a_{EI}(\mathbf{x},i;\mathbf{X},\mathbf{Y},\theta) = \mathbb{E}\left[\max\{0,y_{best}-f(\mathbf{x},i;\mathbf{X},\mathbf{Y},\theta)\}\right]$. Since $f(\mathbf{x},i;\mathbf{X},\mathbf{Y},\theta)\sim \mathcal{N}(\mu(\mathbf{x},i;\mathbf{X},\mathbf{Y},\theta),\Sigma(\mathbf{x},i,\mathbf{x},i;\mathbf{X},\mathbf{Y},\theta))$, the expectation can be computed analytically as follows
\begin{align}
    a_{EI}(\mathbf{x},i;\mathbf{X},\mathbf{Y},\theta) =\varphi(\eta(\mathbf{x},i))) +\sqrt{\Sigma(\mathbf{x},i,\mathbf{x},i;\mathbf{X},\mathbf{Y},\theta)}(\eta(\mathbf{x},i)\Phi(\eta(\mathbf{x},i)),
\end{align}
\begin{align}
    \eta(\mathbf{x},i) & = \frac{y_{best}-\mu(\mathbf{x},i;\mathbf{X},\mathbf{Y},\theta)}{\sqrt{\Sigma(\mathbf{x},i,\mathbf{x},i;\mathbf{X},\mathbf{Y},\theta)}},
\end{align}
where $\Phi(\cdot)$ and $\varphi(\cdot)$ are a cumulative distribution function and a probability density function of the standard normal distribution, respectively. We can choose the next query point by optimize the acquisition function corresponding to all tasks $(\mathbf{x}^{next},i^{next}) = \arg\max_{(\mathbf{x},i)} a_{EI}(\mathbf{x},i;\mathbf{X},\mathbf{Y},\theta)$, or take turns fixing tasks to optimize a single acquisition function.

\subsection{MTBO for Multiple SVM Classifiers}
Support Vector Machines (SVMs) with radial basis kernels are often used for classification tasks, but their performance is susceptible to hyperparameters. Hyperparameter-tuning is a necessary prerequisite for obtaining a good SVM classifier. For multiple SVM classifiers, optimizing their hyperparameters individually requires a lot of repetitive work, especially when their tasks are highly related. With the help of MTBO, the correlation between tasks can be effectively utilized to avoid unnecessary searches and improve the overall optimization speed.
\begin{algorithm}[tb]
  \caption{MTBO for Multiple SVM Classifiers}
  \label{alg:MTBO}
\begin{algorithmic}
  \STATE {\bfseries Input:} $M$ datasets ${\mathcal{D}_1,\mathcal{D}_2,\dots,\mathcal{D}_M}$
  \STATE {\bfseries Parameters:} Positive integers $Iter_1$, $Iter_2$, and $k$
  \STATE {\bfseries Output:} Optimal loss $loss_t^{opt}$ and corresponding hyperparameters $(C_t^{opt},\gamma_t^{opt})$ for each SVM classifier
  \STATE Initialize $\Omega=\phi$
  \FOR{$t=1$ {\bfseries to} $M$}
  \STATE $(C_t^{(1)},\gamma_t^{(1)}) = (1,1)$
  \STATE $loss_t^{(1)} = \mathcal{L}^k(SVM((C_t^{(1)},\gamma_t^{(1)});\mathcal{D}_t))$
  \STATE $\Omega = \Omega \cup \{(C_t^{(1)},\gamma_t^{(1)},loss_t^{(1)}\}$
  \ENDFOR
  \FOR{$i=1$ {\bfseries to} $Iter_1-1$}
  \STATE Select $(C_1^{(i+1)},\gamma_1^{(i+1)})$ by Single-Task Bayesian Optimization.
  \FOR{$t=1$ {\bfseries to} $M$}
  \STATE $loss_t^{(i+1)} = \mathcal{L}^k(SVM((C_t^{(i+1)},\gamma_t^{(i+1)});\mathcal{D}_t))$
  \STATE $(C_t^{(i+1)},\gamma_t^{(i+1)}) = (C_1^{(i+1)},\gamma_1^{(i+1)})$
  \STATE $\Omega = \Omega \cup \{(C_t^{(i+1)},\gamma_t^{(i+1)},loss_t^{(i+1)}\}$
  \ENDFOR
  \ENDFOR
  \STATE Fitting a Multi-Task Gaussian Process based on $\Omega$.
  \FOR{$i=Iter_1$ {\bfseries to} $Iter_2$}
  \STATE Fix the task index: $t=\mod (i-Iter_1,M)+1$
  \STATE Determine the current iteration time for the $t^{th}$ task: $j=Iter_1+1+(i-Iter_1-t+1)/M$
  \STATE $(C_t^{(j)},\gamma_t^{(j)}) = \arg \max_{(C,\gamma)} a_{EI}(C,\gamma,t;\Omega,\theta)$
  \STATE $loss_t^{(j)} = \mathcal{L}^k(SVM((C_t^{(j)},\gamma_t^{(j)});\mathcal{D}_t))$
  \STATE $\Omega = \Omega \cup \{(C_t^{(j)},\gamma_t^{(j)},loss_t^{(j)}\}$
  \STATE Update the Multi-Task Gaussian Process
  \ENDFOR
  \FOR{$t=1$ \bfseries to $M$}
  \STATE $i^* = \arg \min_i loss_t^{(i)}$
  \STATE $loss_t^{opt} = loss_t^{(i^*)}$, $(C_t^{opt},\gamma_t^{opt}) = (C_t^{(i^*)},\gamma_t^{(i^*)})$
  \ENDFOR
\end{algorithmic}
\end{algorithm}
The procedure of MTBO for multiple SVM classifier is shown in pseudocode in \cref{alg:MTBO}. ${\mathcal{D}_1,\mathcal{D}_2,\dots,\mathcal{D}_M}$ are the given datasets used to train the SVM classifier. In our work, we generated nine datasets through varies discretization strategies. Parameters ${Iter_1}$, ${Iter_2}$, and ${k}$ need to be given in advance. $Iter_1$ should take a suitable value so that there are enough observations to estimate the parameters, and without wasting too much time to train the SVMs. $Iter_1$ and $Iter_2$ determine the total number of iterations. $\mathcal{L}^k(SVM((C_t^{(i)},\gamma_t^{(i)});\mathcal{D}_t))$ represents the loss of $k$-fold cross-validation on dataset $\mathcal{D}_t$ with $(C_t^{(i)},\gamma_t^{(i)})$ as the hyperparameters of the SVM.

\section{Experiments}

\subsection{Data Description and Preprocessing}
The in-house dataset we used in this work consists of 499 patients with pathologically confirmed pulmonary nodules, treated between September 2005 and August 2013. All patients received whole body CT scans for diagnosis.

\begin{figure}[htbp]
\begin{center}
\centerline{\includegraphics[width=.8\columnwidth]{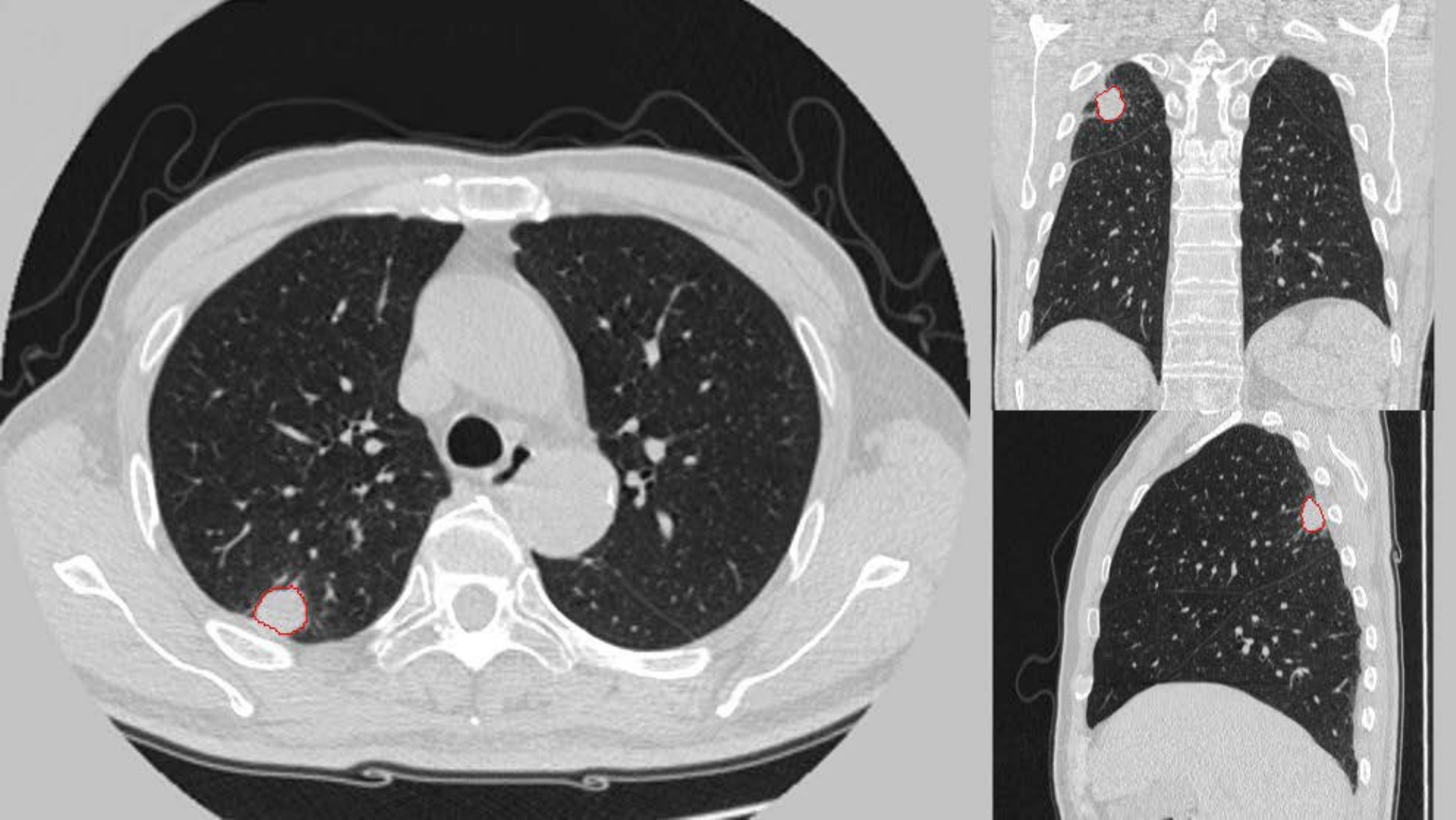}}
\caption{\textbf{Segmentation results under different views}. Left: axial; top right: coronal; bottom right: sagittal.}
\label{fig:seg}
\end{center}
\end{figure}

The nodule segmentation is performed semi-automatically, that is, the radiologist manually locates the nodule and examines the final segmentation results, and the contour of the nodule is automatically delineated by the computer programs. First, a trained U-net (R-231) \cite{hofmanninger2020} was employed to remove regions outside the lungs in CT, thereby separating the juxta-pleural nodules from the lung boundary. Second, all images were resampled to 1×1×1-mm voxel size using cubic interpolation. Third, the boundary of nodule was described by Active contour model \cite{chan2001} and refined by morphological methods. Finally, we corrected the segmentation based on the radiologist's feedback on the results. \cref{fig:seg} shows an example of segmentation results in three views.

\subsection{Radiomic Features Extraction}

Radiomics, a technique for quantitatively extracting image features from medical images to describe tumor phenotypes, is gaining importance in cancer imaging \cite{Yip2016,gillies2016,Mayerhoefer2020}. We refer to \cite{park2020} to extract a series of radiomic features from each ROI, including shape features (n=8), first-order statistics features (n=13), and second-order statistics features derived from gray-level co-occurrence matrix (GLCM; n=23) and gray-level run-length matrix (GLRLM; n=12). The details of the features are listed in \cref{tab:features}. Since we generate nine discretization strategies, correspondingly we can get nine sets of feature matrices. 

\begin{table}[htbp]
\caption{Radiomic features extracted in this study.}
\label{tab:features}
\centering
\begin{tabular}{lcp{10cm}}
\toprule
Type & No. & Feature Names \\
\midrule
Shape    & 8& Spherical disproportion, sphericity, surface area, surface-area-to-volume ratio, volume, equivalent diameter, extent, and solidity \\
1st-order & 13 & Mean, var, skewness, kurtosis, median, min, mad, max, range, cov, rms, entropy, and uniformity\\
GLCM & 23 & Autocorrelation, cluster prominence, cluster shade, cluster tendency, contrast, correlation, difference entropy, difference variance, dissimilarity, energy, entropy, homogeneity, informational measure of correlation 1 \& 2, inverse difference moment, inverse difference moment normalized, inverse difference normalized, inverse variance, maximum probability, mean, sum entropy, sum variance, and variance\\
GLRLM & 12 & GLN, HGRE, LRE, LRHGE, LRLGE, LGRE, N\_runs, RLN, RP, SRE, SRHGE, and SRLGE\\
\bottomrule
\end{tabular}
\end{table}

\subsection{Classification and Evaluation}

Nine SVM classifiers with radial basis function kernels are trained for distinguishing the benign-malignant pulmonary nodules. The goal is to minimize the classification loss obtained by $k$-fold cross-validation for each SVM. Penalty factor $C$  and kernel size $\gamma$  are two hyperparameters that need to be optimized. We restrict the hyperparameters $C$ and $\gamma$ to $[10^{-3},10^{3}]$. In addition, since the value range of the loss function $\mathcal{L}(\cdot)$ is $[0,1]$, we use a monotonic transformation to map the loss function onto $\mathbb{\Bar{R}}$, so that the transformed loss functions can be fitted with Gaussian process. The transformed loss functions can be written as:
\begin{align}\label{eq:tan}
    f(C,\gamma,i) = \tan (\pi\cdot \mathcal{L}^k(C,\gamma,i)-\pi/2),\\ \nonumber i=1,2,\dots,9.
\end{align}
Then, the optimal hyperparameters can be searched for each classifier using multi-task Bayesian optimization. Single-Task Bayesian Optimization (STBO) is employed to solve each objective function separately as baseline.

To measure the fitting effect of the single/multi-task Gaussian process, we compute the root mean squared error (RMSE):
\begin{align}
    RMSE^i = &\sqrt{\frac{1}{N_1N_2}\sum_{j=0}^{N_1} \sum_{k=0}^{N_2} \left(f(C_j,\gamma_k,i)  - \hat{f}(C_j,\gamma_k,i)\right)^2}, \nonumber \\  &i  = 1,2,\dots,9.
\end{align}
we can easily recalculate the original loss according to the transformed loss, and the same is true for RMSE. In practice, we set $N_1=N_2=60$, and $C_j=\gamma_j=10^{-3+j/10}$. 

\subsection{Result and Discussion}
For STBO, we use $(C^{(1)},\gamma^{(1)}) = (1,1)$ as the initial point to determine the next query by maximizing the EI acquisition function, each task iterate 30 times; for MTBO, we use \cref{alg:MTBO} by setting $Iter_1=10$, $Iter_2=190$, and $k=10$. Specifically, we first adopt STBO, also starting from $(1, 1)$, and iteratively search on the $1^{st}$ task for 10 times. Other tasks are evaluated on these same 10 queries. Thus, we obtain 90 complete observations to fit a multi-task Gaussian process. Then, we take turns fixing the tasks for a total of 180 iterative searches, thus ensuring that each task is also evaluated 30 times. \cref{tab:result} shows the optimal hyperparameters for each task obtained with STBO and MTBO, respectively, and the classification loss under 10-fold cross-validation by training SVM classifiers with the optimal hyperparameters.
\begin{table}[htbp]
\caption{Optimization results for each task.}
\label{tab:result}
\centering
\begin{tabular}{lrrrr}
\toprule
 \makecell[c]{Task} & \makecell[c]{STBO \\ Hyperparameters} &\makecell[c]{STBO \\ Loss} & \makecell[c]{MTBO \\ Hyperparameters} &\makecell[c]{MTBO \\ Loss}  \\
\midrule
N=16, min-max & (96.3927,137.1610) & \textbf{0.1884} &(6.7267,27.6362)&\textbf{0.1884}\\
N=16, mean ± 2SD & (1.1624,10.7750) &0.1904&(24.2547,35.4013)&\textbf{0.1864}\\
N=16, mean ± 3SD & (1000.0000,246.5459) &\textbf{0.1844}&(20.1826,31.5285)&0.1864\\
N=32, min-max & (19.8362,50.2647) &\textbf{0.1844}&369.7450,245.7809&\textbf{0.1844}\\
N=32, mean ± 2SD & (286.2686,63.2712) &\textbf{0.1824}&(11.9257,24.7766)&0.1844\\
N=32, mean ± 3SD & (879.7315,237.3126) &\textbf{0.1964}&(369.7450,245.7809)&0.1984\\
N=64, min-max & (15.1667,20.4187) &\textbf{0.1804}&(23.1439,34.3956)&\textbf{0.1804}\\
N=64, mean ± 2SD & (103.7646,61.2323) &0.1884&(48.9185,53.5978)&\textbf{0.1864}\\
N=64, mean ± 3SD & (264.1003,43.5839) &0.1944&(16.2418,32.1168)&\textbf{0.1824}\\
\bottomrule
\end{tabular}
\end{table}

\begin{figure*}[htbp]
\begin{center}
\centerline{\includegraphics[width=\textwidth]{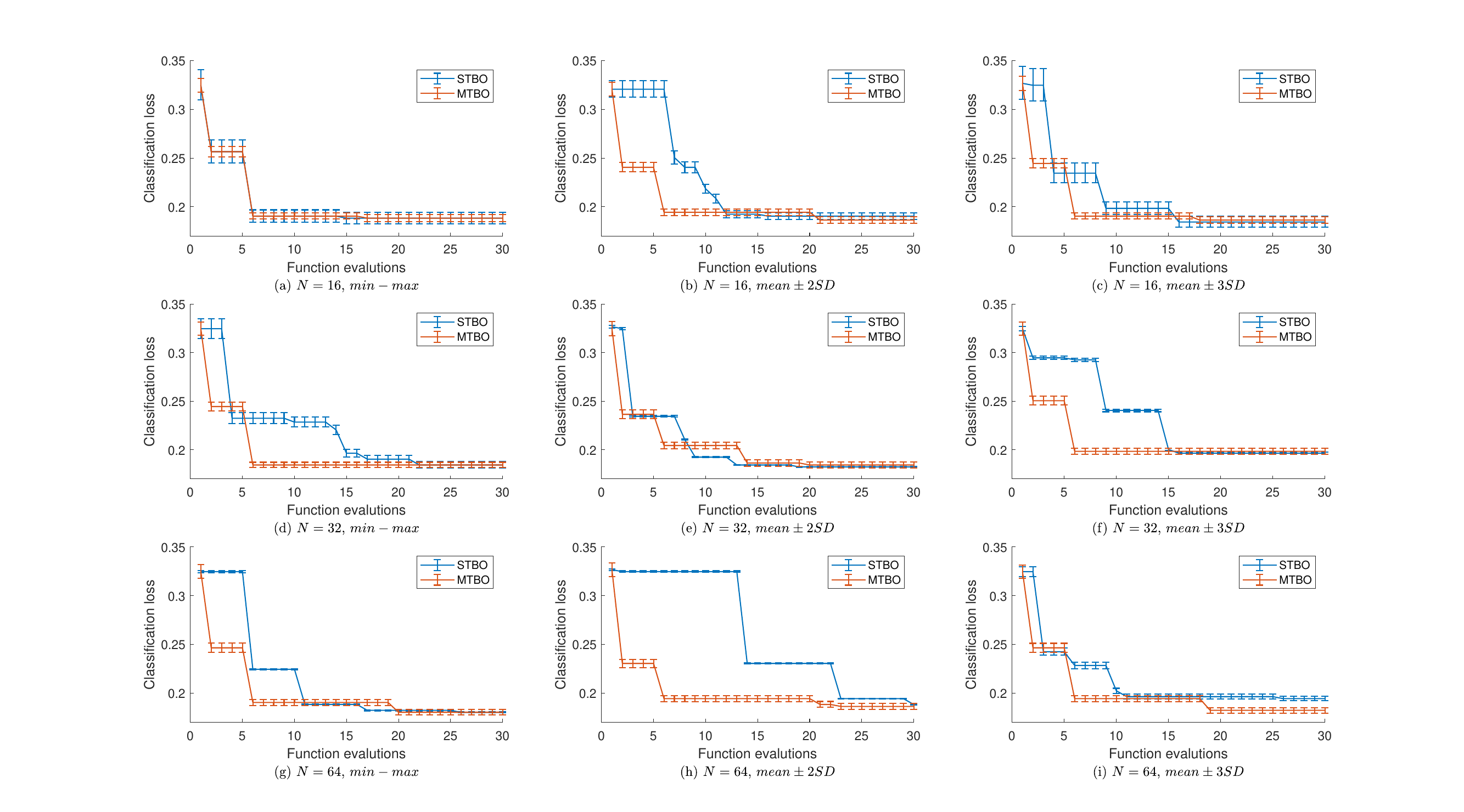}}
\caption{\textbf{Classification loss obtained by 10-fold cross-validation using STBO and MTBO respectively.}}
\label{fig:result}
\end{center}
\end{figure*}

To better understand the behavior of the different methods, we plot the number of evaluations versus the current optimal loss in \cref{fig:result}. In the first ten evaluations, we found that even if the queries of the first task are directly transferred to other tasks, good performance can be achieved. This also shows that the nine tasks we need to optimize are highly related. As shown in \cref{tab:result} and \cref{fig:result}, after 30 iterations, there is not much difference between MTBO and STBO in finding the minimum loss for each task, and the optimal hyperparameters may not unique. It is worth mentioning that although there is no difference in the final result, it can be seen from \cref{fig:result} that MTBO always faster to reaches the local optimum. In practical applications, this result makes us more inclined to choose MTBO for hyperparameter-tuning.

\begin{figure}[htbp]
    \centering
    \subfigure[Original]{
    \label{fig:a}
    \includegraphics[width=0.3\columnwidth]{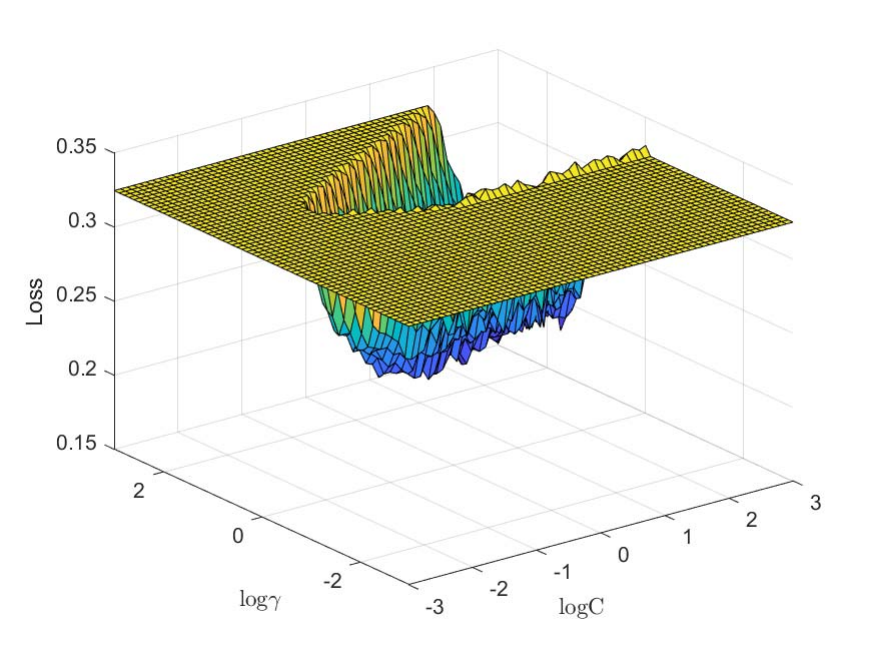}
    }
    \subfigure[STBO]{
    \label{fig:b}
    \includegraphics[width=0.3\columnwidth]{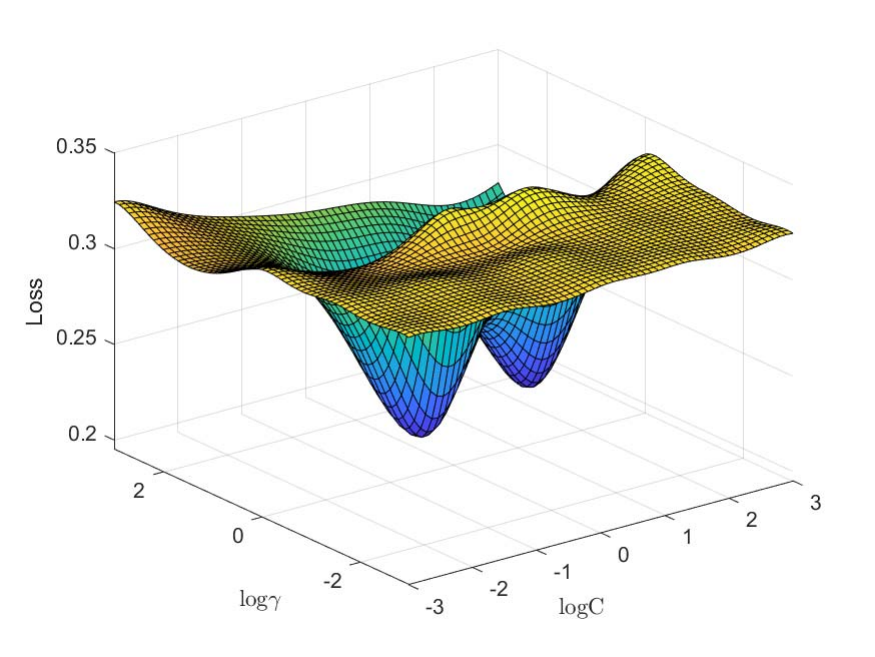}
    }
    \subfigure[MTBO]{
    \label{fig:c}
    \includegraphics[width=0.3\columnwidth]{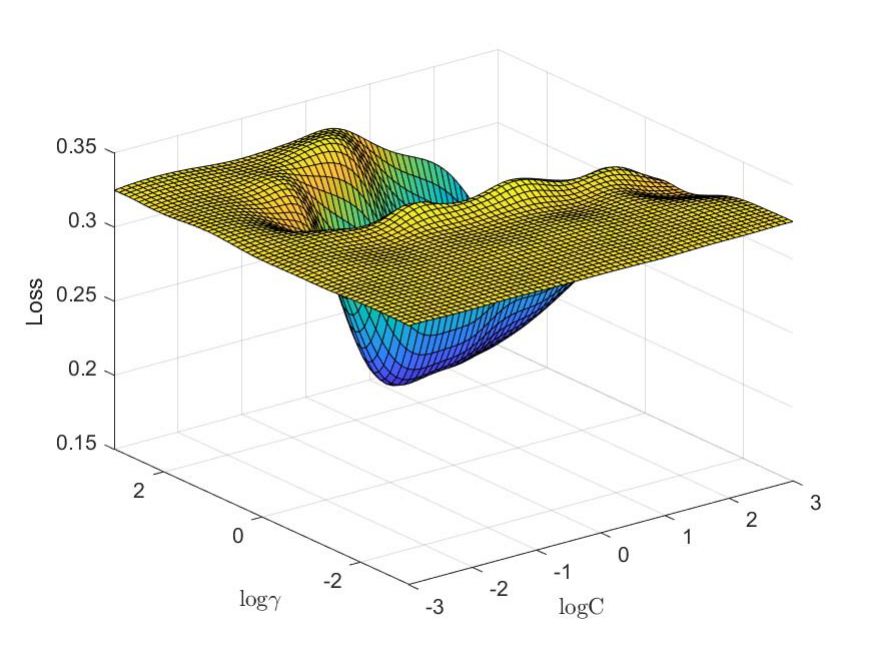}
    }
    \caption{\textbf{The landscape for loss function on hyperparameter space.} (N = 64, mean ± 3SD)}
    \label{fig:enter-label}
\end{figure}

After the above experiments, we evaluate the loss for each task on over 3600 sets of hyperparameters. Taking the task {N=64,mean ± 3SD} as an example, \cref{fig:a} intuitively shows the relationship between the loss and hyperparameters, and \cref{fig:b,fig:c} are the surfaces fitted for the task by the single/multi-task Gaussian process, respectively. By comparing these three figures, it can be found that the surface fitted by the multi-task Gaussian process is closer to the real surface than single-task Gaussian process, and the results in \cref{tab:rmse} also confirm this. At the same time, we find that the prediction variance during single-task Bayesian optimization is very small (see \cref{fig:result} (e)-(i)). Combining the above findings, it shows that the single-task Gaussian process is slightly overfitting. While the fitting of the multi-task Gaussian process utilizes the information of the related-tasks. With more data, more robust and general representations can be learned for multiple tasks, resulting in better sharing of knowledge between tasks and less risk of overfitting in each task.

\begin{table}[htbp]
\caption{RMSE obtained by single/multi-task Gaussian Process.}
\label{tab:rmse}
\centering
\begin{tabular}{lrr}
\toprule
 \makecell[c]{Task} & \makecell[c]{Single-task GP \\ RMSE} & \makecell[c]{Multi-task GP \\ RMSE} \\
\midrule
N=16, min-max & 0.0311 & \textbf{0.0141} \\
N=16, mean ± 2SD & 0.0266 & \textbf{0.0124}\\
N=16, mean ± 3SD & 0.0256 & \textbf{0.0109}\\
N=32, min-max & 0.0273 & \textbf{0.0108} \\
N=32, mean ± 2SD & 0.0283 & \textbf{0.0093}\\
N=32, mean ± 3SD & 0.0161 & \textbf{0.0093}\\
N=64, min-max & 0.0206 & \textbf{0.0105}\\
N=64, mean ± 2SD & 0.0301 & \textbf{0.0092}\\
N=64, mean ± 3SD & 0.0199 & \textbf{0.0097}\\
\bottomrule
\end{tabular}
\end{table}

In addition, we found that there is no significant difference in model performance under each discretization strategy, so no discretization strategy can be considered optimal only from the perspective of classification loss. Considering the robustness of the strategy, using mean ± 2SD or mean ± 3SD as the quantization range helps to resist noise caused by segmentation or other reasons, such as removing high-density bones in CT. Moreover, choosing a smaller bin number helps to keep the results consistent across multiple imaging and has a significant computational complexity advantage in the feature extraction stage. Therefore, in practice we may prefer to choose a smaller bin number and a more robust quantization range as the image discretization strategy.

\section{Conclusion}
Medical image analysis based on machine learning currently still face daunting challenges. Our research first introduce the MTBO to the medical imaging and efficiently found satisfactory hyperparameters for RBF SVM classifiers in the face of multiple diagnostic tasks. In addition to image discretization, there are a large number of other non-standardized strategies in the medical image analysis, including filtering, de-noising, and segmentation. This research provides an idea for how to quickly evaluate each strategy. It should be noted that this study has some limitations.
\begin{itemize}
    \item \textit{Time-saving:} In this study, training a SVM classifier is not time-consuming, instead it takes more time to fit the data using a multi-task Gaussian process. This makes it more efficient to process multiple tasks one by one using single-task Bayesian optimization. The advantages of MTBO will appear when the step of evaluation is really time-consuming, such as training a deep neural network. In this case, the time required to fit a multi-task Gaussian process is negligible.
    \item \textit{Multitasking:} When we are faced with more tasks, it may be inefficient to directly apply our method, because computational time and memory for large scale matrix operations such as inversion cannot be ignored. In this case some inexact method is needed to speed up multi-task Bayesian optimization.
    \item \textit{Cold starting:} MTBO requires good complete observations of a certain scale to give appropriate initial estimates of model parameters. This paper gives a method for how to choose these initial points to start the model, but no in-depth study of how many observations to choose.
\end{itemize}
\section*{Funding}
This study was supported by the Science and Technology Planning Project of Guangzhou City (202201020452).
\bibliographystyle{unsrt}  
\bibliography{references}

@ARTICLE{chan2001,
  author={Chan, T.F. and Vese, L.A.},
  journal={IEEE Transactions on Image Processing}, 
  title={Active contours without edges}, 
  year={2001},
  volume={10},
  number={2},
  pages={266-277},
}

@article{Bergstra2012,
  author  = {James Bergstra and Yoshua Bengio},
  title   = {Random Search for Hyper-Parameter Optimization},
  journal = {Journal of Machine Learning Research},
  year    = {2012},
  volume  = {13},
  number  = {10},
  pages   = {281-305},
}

@inproceedings{Snoek2012,
 author = {Snoek, Jasper and Larochelle, Hugo and Adams, Ryan P},
 booktitle = {Advances in Neural Information Processing Systems},
 pages = {2951-2959},
 publisher = {Curran Associates, Inc.},
 title = {Practical {Bayesian} Optimization of Machine Learning Algorithms},
 volume = {25},
 year = {2012},
}

@INPROCEEDINGS{Borgli2019,
  author={Borgli, Rune Johan and Kvale Stensland, Håkon and Riegler, Michael Alexander and Halvorsen, Pål},
  booktitle={2019 13th International Symposium on Medical Information and Communication Technology (ISMICT)}, 
  title={Automatic Hyperparameter Optimization for Transfer Learning on Medical Image Datasets Using {Bayesian} Optimization}, 
  year={2019},
  volume={},
  number={},
  pages={1-6},
  publisher = {IEEE},
}

@article{gevaert2012,
author = {Gevaert, Olivier and Xu, Jiajing and Hoang, Chuong D. and Leung, Ann N. and Xu, Yue and Quon, Andrew and Rubin, Daniel L. and Napel, Sandy and Plevritis, Sylvia K.},
title = {Non-Small Cell Lung Cancer: Identifying Prognostic Imaging Biomarkers by Leveraging Public Gene Expression Microarray Data-Methods and Preliminary Results},
journal = {Radiology},
volume = {264},
number = {2},
pages = {387-396},
year = {2012},
}

@article{huang2016,
author = {Huang, Yan-qi and Liang, Chang-hong and He, Lan and Tian, Jie and Liang, Cui-shan and Chen, Xin and Ma, Ze-lan and Liu, Zai-yi},
title = {Development and Validation of a Radiomics Nomogram for Preoperative Prediction of Lymph Node Metastasis in Colorectal Cancer},
journal = {Journal of Clinical Oncology},
volume = {34},
number = {18},
pages = {2157-2164},
year = {2016},
}

@inproceedings{swersky2013,
 author = {Swersky, Kevin and Snoek, Jasper and Adams, Ryan P},
 booktitle = {Advances in Neural Information Processing Systems},
 pages = {2004-2012},
 publisher = {Curran Associates, Inc.},
 title = {Multi-Task {Bayesian} Optimization},
 volume = {26},
 year = {2013},
}

@article{Yip2016,
	year = {2016},
	publisher = {{IOP} Publishing},
	volume = {61},
	number = {13},
	pages = {R150--R166},
	author = {Stephen S F Yip and Hugo J W L Aerts},
	title = {Applications and limitations of radiomics},
	journal = {Physics in Medicine and Biology},
}

@article{leijenaar2013,
author = {Ralph T. H. Leijenaar and Sara Carvalho and Emmanuel Rios Velazquez and Wouter J. C. van Elmpt and Chintan Parmar and Otto S. Hoekstra and Corneline J. Hoekstra and Ronald Boellaard and André L. A. J. Dekker and Robert J. Gillies and Hugo J. W. L. Aerts and Philippe Lambin},
title = {Stability of {FDG-PET} Radiomics features: An integrated analysis of test-retest and inter-observer variability},
journal = {Acta Oncologica},
volume = {52},
number = {7},
pages = {1391-1397},
year  = {2013},
publisher = {Taylor & Francis},
}

@article{shafiq2018,
  title={Voxel size and gray level normalization of {CT} radiomic features in lung cancer},
  author={Shafiq-ul-Hassan, Muhammad and Latifi, Kujtim and Zhang, Geoffrey and Ullah, Ghanim and Gillies, Robert and Moros, Eduardo},
  journal={Scientific reports},
  volume={8},
  number={1},
  pages={1-9},
  year={2018},
  publisher={Nature Publishing Group},
}

@article{park2020,
  title={Reliability of {CT} radiomic features reflecting tumour heterogeneity according to image quality and image processing parameters},
  author={Park, Bum Woo and Kim, Jeong Kon and Heo, Changhoe and Park, Kye Jin},
  journal={Scientific reports},
  volume={10},
  number={1},
  pages={1-13},
  year={2020},
  publisher={Nature Publishing Group},
}

@article{larue2017,
author = {Ruben T. H. M. Larue and Janna E. van Timmeren and Evelyn E. C. de Jong and Giacomo Feliciani and Ralph T. H. Leijenaar and Wendy M. J. Schreurs and Meindert N. Sosef and Frank H. P. J. Raat and Frans H. R. van der Zande and Marco Das and Wouter van Elmpt and Philippe Lambin},
title = {Influence of gray level discretization on radiomic feature stability for different {CT} scanners, tube currents and slice thicknesses: a comprehensive phantom study},
journal = {Acta Oncologica},
volume = {56},
number = {11},
pages = {1544-1553},
year  = {2017},
publisher = {Taylor & Francis},
}

@ARTICLE{shahriari2016,
  author={Shahriari, Bobak and Swersky, Kevin and Wang, Ziyu and Adams, Ryan P. and de Freitas, Nando},
  journal={Proceedings of the IEEE}, 
  title={Taking the Human Out of the Loop: A Review of {Bayesian} Optimization}, 
  year={2016},
  volume={104},
  number={1},
  pages={148-175},
}

@INPROCEEDINGS{khosravan2018,
  author={Khosravan, Naji and Bagci, Ulas},
  booktitle={2018 40th Annual International Conference of the IEEE Engineering in Medicine and Biology Society (EMBC)}, 
  title={Semi-Supervised Multi-Task Learning for Lung Cancer Diagnosis}, 
  year={2018},
  volume={},
  number={},
  pages={710-713},
}

@INPROCEEDINGS{Bonilla2007,
 author = {Bonilla, Edwin V and Chai, Kian and Williams, Christopher},
 booktitle = {Advances in Neural Information Processing Systems},
 pages = {153-160},
 publisher = {Curran Associates, Inc.},
 title = {Multi-task {Gaussian} Process Prediction},
 volume = {20},
 year = {2007},
}

@article{gonen2014,
    author = {Gönen, Mehmet and Margolin, Adam A.},
    title = {Drug susceptibility prediction against a panel of drugs using kernelized {Bayesian} multitask learning},
    journal = {Bioinformatics},
    volume = {30},
    number = {17},
    pages = {i556-i563},
    year = {2014},
}

@article{caruana1997,
Author = {Caruana, R},
Title = {Multitask learning},
Journal = {MACHINE LEARNING},
Year = {1997},
Volume = {28},
Number = {1},
Pages = {41-75},
}

@InProceedings{chandra2016,
author="Chandra, Rohitash
and Gupta, Abhishek
and Ong, Yew-Soon
and Goh, Chi-Keong",
title="Evolutionary Multi-task Learning for Modular Training of Feedforward Neural Networks",
booktitle="Neural Information Processing",
year="2016",
publisher="Springer International Publishing",
address="Cham",
pages="37-46",
}

@article{hawkins2014,
  title={Predicting outcomes of nonsmall cell lung cancer using {CT} image features},
  author={Hawkins, Samuel H and Korecki, John N and Balagurunathan, Yoganand and Gu, Yuhua and Kumar, Virendra and Basu, Satrajit and Hall, Lawrence O and Goldgof, Dmitry B and Gatenby, Robert A and Gillies, Robert J},
  journal={IEEE access},
  volume={2},
  pages={1418-1426},
  year={2014},
  publisher={IEEE},
}

@article{junior2018,
  title={Radiomics-based features for pattern recognition of lung cancer histopathology and metastases},
  author={Junior, Jos{\'e} Raniery Ferreira and Koenigkam-Santos, Marcel and Cipriano, Federico Enrique Garcia and Fabro, Alexandre Todorovic and de Azevedo-Marques, Paulo Mazzoncini},
  journal={Computer methods and programs in biomedicine},
  volume={159},
  pages={23-30},
  year={2018},
  publisher={Elsevier},
}

@inproceedings{katre2017,
  title={Detection of lung cancer stages using image processing and data classification techniques},
  author={Katre, Pooja R and Thakare, Anuradha},
  booktitle={2017 2nd International Conference for Convergence in Technology (I2CT)},
  pages={402-404},
  year={2017},
  organization={IEEE},
}

@inproceedings{kulkarni2014,
  title={Classification of lung cancer stages on {CT} scan images using image processing},
  author={Kulkarni, Anjali and Panditrao, Anagha},
  booktitle={2014 IEEE International Conference on Advanced Communications, Control and Computing Technologies},
  pages={1384-1388},
  year={2014},
  organization={IEEE},
}

@article{liu2020,
  title={Evolving the pulmonary nodules diagnosis from classical approaches to deep learning-aided decision support: three decades’ development course and future prospect},
  author={Liu, Bo and Chi, Wenhao and Li, Xinran and Li, Peng and Liang, Wenhua and Liu, Haiping and Wang, Wei and He, Jianxing},
  journal={Journal of cancer research and clinical oncology},
  volume={146},
  number={1},
  pages={153-185},
  year={2020},
  publisher={Springer},
}

@article{shen2019,
  title={An interpretable deep hierarchical semantic convolutional neural network for lung nodule malignancy classification},
  author={Shen, Shiwen and Han, Simon X and Aberle, Denise R and Bui, Alex A and Hsu, William},
  journal={Expert systems with applications},
  volume={128},
  pages={84-95},
  year={2019},
  publisher={Elsevier},
}

@article{shi2005,
  title={Tumor classification by tissue microarray profiling: random forest clustering applied to renal cell carcinoma},
  author={Shi, Tao and Seligson, David and Belldegrun, Arie S and Palotie, Aarno and Horvath, Steve},
  journal={Modern Pathology},
  volume={18},
  number={4},
  pages={547-557},
  year={2005},
  publisher={Nature Publishing Group},
}

@article{sun2013,
  title={Comparative evaluation of support vector machines for computer aided diagnosis of lung cancer in {CT} based on a multi-dimensional data set},
  author={Sun, Tao and Wang, Jingjing and Li, Xia and Lv, Pingxin and Liu, Fen and Luo, Yanxia and Gao, Qi and Zhu, Huiping and Guo, Xiuhua},
  journal={Computer methods and programs in biomedicine},
  volume={111},
  number={2},
  pages={519-524},
  year={2013},
  publisher={Elsevier},
}

@article{touw2013,
  title={Data mining in the Life Sciences with Random Forest: a walk in the park or lost in the jungle?},
  author={Touw, Wouter G and Bayjanov, Jumamurat R and Overmars, Lex and Backus, Lennart and Boekhorst, Jos and Wels, Michiel and van Hijum, Sacha AFT},
  journal={Briefings in bioinformatics},
  volume={14},
  number={3},
  pages={315-326},
  year={2013},
  publisher={Oxford University Press},
}

@article{xie2018,
  title={Knowledge-based collaborative deep learning for benign-malignant lung nodule classification on chest {CT}},
  author={Xie, Yutong and Xia, Yong and Zhang, Jianpeng and Song, Yang and Feng, Dagan and Fulham, Michael and Cai, Weidong},
  journal={IEEE transactions on medical imaging},
  volume={38},
  number={4},
  pages={991-1004},
  year={2018},
  publisher={IEEE},
}

@article{yuan2018,
  title={Hybrid-feature-guided lung nodule type classification on {CT} images},
  author={Yuan, Jingjing and Liu, Xinglong and Hou, Fei and Qin, Hong and Hao, Aimin},
  journal={Computers \& Graphics},
  volume={70},
  pages={288-299},
  year={2018},
  publisher={Elsevier},
}

@article{zhou2018,
  title={Diagnosis of distant metastasis of lung cancer: based on clinical and radiomic features},
  author={Zhou, Hongyu and Dong, Di and Chen, Bojiang and Fang, Mengjie and Cheng, Yue and Gan, Yuncun and Zhang, Rui and Zhang, Liwen and Zang, Yali and Liu, Zhenyu and others},
  journal={Translational oncology},
  volume={11},
  number={1},
  pages={31-36},
  year={2018},
  publisher={Elsevier},
}

@article{hofmanninger2020,
author = {Hofmanninger, Johannes and Prayer, Forian and Pan, Jeanny and Röhrich, Sebastian and Prosch, Helmut and Langs, Georg},
year = {2020},
pages = {50},
title = {Automatic lung segmentation in routine imaging is primarily a data diversity problem, not a methodology problem},
volume = {4},
journal = {European Radiology Experimental},
}

@article {Mayerhoefer2020,
	author = {Mayerhoefer, Marius E. and Materka, Andrzej and Langs, Georg and H{\"a}ggstr{\"o}m, Ida and Szczypi{\'n}ski, Piotr and Gibbs, Peter and Cook, Gary},
	title = {Introduction to Radiomics},
	volume = {61},
	number = {4},
	pages = {488--495},
	year = {2020},
	publisher = {Society of Nuclear Medicine},
	journal = {Journal of Nuclear Medicine},
}

@article{gillies2016,
author = {Gillies, Robert J. and Kinahan, Paul E. and Hricak, Hedvig},
title = {Radiomics: Images Are More than Pictures, They Are Data},
journal = {Radiology},
volume = {278},
number = {2},
pages = {563-577},
year = {2016},
}

@article{chatterjeeadvancement,
  title={Advancement of Machine Learning Studies on Lung Cancer (A literature survey from year 2015 to 2020)},
  author={Chatterjee, Pratik and Paul, Utpalendu and Gupta, Nandini},
  volume = {9},
  number = {1},
  year = {2021},
  journal = {International Journal of All Research Education and Scientific Methods},
  pages = {370-381},
}

@article{travis2013new,
  title={New pathologic classification of lung cancer: relevance for clinical practice and clinical trials},
  author={Travis, William D and Brambilla, Elisabeth and Riely, Gregory J},
  journal={J Clin Oncol},
  volume={31},
  number={8},
  pages={992--1001},
  year={2013},
}

@article{li2019performance,
  title={The performance of deep learning algorithms on automatic pulmonary nodule detection and classification tested on different datasets that are not derived from LIDC-IDRI: a systematic review},
  author={Li, Dana and Mikela Vilmun, Bolette and Frederik Carlsen, Jonathan and Albrecht-Beste, Elisabeth and Ammitzb{\o}l Lauridsen, Carsten and Bachmann Nielsen, Michael and Lindskov Hansen, Kristoffer},
  journal={Diagnostics},
  volume={9},
  number={4},
  pages={207},
  year={2019},
  publisher={Multidisciplinary Digital Publishing Institute},
}

@article{van2016repeatability,
  title={Repeatability of radiomic features in non-small-cell lung cancer {[18 F] FDG-PET/CT} studies: impact of reconstruction and delineation},
  author={van Velden, Floris HP and Kramer, Gerbrand M and Frings, Virginie and Nissen, Ida A and Mulder, Emma R and de Langen, Adrianus J and Hoekstra, Otto S and Smit, Egbert F and Boellaard, Ronald},
  journal={Molecular imaging and biology},
  volume={18},
  number={5},
  pages={788--795},
  year={2016},
  publisher={Springer},
}

\end{document}